\documentclass[a4paper]{article}

\usepackage{INTERSPEECH2021}

\title{MTI-Net: A Multi-Target Speech Intelligibility Prediction Model}
\name{Ryandhimas E. Zezario$^1$$^2$, Szu-wei Fu,$^3$, Fei Chen$^4$, Chiou-Shann Fuh$^1$, Hsin-Min Wang$^2$, Yu Tsao$^2$}
\address{
  $^1$National Taiwan University,
  $^2$Academia Sinica,
  $^3$Microsoft Corporation,\\
  $^4$ Southern University of Science and Technology of China}
\email{\{ryandhimas, yu.tsao\}@citi.sinica.edu.tw}

\begin{document}

\maketitle
\begin{abstract}
Recently, deep learning (DL)-based non-intrusive speech assessment models have attracted great attention. Many studies report that these DL-based models yield satisfactory assessment performance and good flexibility, but their performance in unseen environments remains a challenge. Furthermore, compared to quality scores, fewer studies elaborate deep learning models to estimate intelligibility scores. This study proposes a multi-task speech intelligibility prediction model, called MTI-Net, for simultaneously predicting human and machine intelligibility measures. Specifically, given a speech utterance, MTI-Net is designed to predict human subjective listening test results and word error rate (WER) scores. We also investigate several methods that can improve the prediction performance of MTI-Net. First, we compare different features (including low-level features and embeddings from self-supervised learning (SSL) models) and prediction targets of MTI-Net. Second, we explore the effect of transfer learning and multi-tasking learning on training MTI-Net. Finally, we examine the potential advantages of fine-tuning SSL embeddings. Experimental results demonstrate the effectiveness of using cross-domain features, multi-task learning, and fine-tuning SSL embeddings. Furthermore, it is confirmed that the intelligibility and WER scores predicted by MTI-Net are highly correlated with the ground-truth scores. 

\end{abstract}
\noindent\textbf{Index Terms}: Subjective listening tests, WER, STOI, speech intelligibility prediction, self-supervised learning

\section{Introduction}
For many speech-related applications, such as hearing aids, telecommunications, and automatic speech recognition (ASR), speech intelligibility is a key metric of user satisfaction. The metric measures the ratio of correctly recognized words to total words in a listening test, in which participants listen to a set of speech and answer what they hear. 
Since the measurements are based on human hearing tests, collecting data from a sufficient number of listeners is critical for unbiased measurements. However, conducting large-scale hearing tests is prohibitive. To overcome this problem, many signal processing-based intelligibility metrics have been proposed as surrogate metrics for human listening behavior, such as articulation index (AI) \cite{ref_35}, speech intelligibility index (SII) \cite{ref_36}, extended SII (ESII) \cite{ref_37}, speech transmission index (STI) \cite{ref_38}, and short-time objective intelligibility (STOI) \cite{ref_39}. These signal processing-based intelligibility metrics can be roughly divided into two categories, intrusive and non-intrusive metrics. Intrusive metrics \cite{ref_35,ref_36,ref_37,ref_38,ref_39} require the corresponding clean speech as a reference to measure intelligibility scores, while non-intrusive metrics do not need a reference \cite{srmr, moda}. Compared to intrusive metrics, non-intrusive metrics have better flexibility but generally provide lower prediction accuracy.  

Recently, deep learning (DL)  models have been introduced into speech intelligibility prediction systems. For these systems, a DL model is used as a regression function to predict the intelligibility score given a speech signal. 
Based on the types of ground-truth labels, these DL-based prediction systems can be divided into two categories: one that predicts objective evaluation scores, such as STI, and STOI \cite{ ref_55, ref_56, ref_52}, and the other that predicts human subjective ratings of human listening tests \cite{andersen2018nonintrusive, SIP_DL,TMINT-QI}. Due to differences in the listening abilities of the listeners, the range of human subjective ratings is larger than that of objective scores, as reported in \cite{TMINT-QI}. Therefore, training an intelligibility  model that predicts human subjective ratings is more difficult than training a model that predicts objective scores. %To facilitate a more effective training process, several approaches have been proposed. For examples, (1) DNSMOS \cite{dnsmos} uses a teacher-student architecture to generate pseudo labels to improve the assessment accuracy, (2) MBNet uses an additional bias network to compensate for differences among listeners \cite{mbnet_mos}, and (3) SSL-MOS \cite{ssl-mos} proposes to fine-tune the self-supervised learning (SSL) model to improve the prediction results. 

%Therefore, we aim to conduct a comprehensive study to achieve a more robust intelligibility prediction system in this study.

In our previous work, we proposed a multi-objective speech assessment model, called MOSA-Net \cite{mosa-net}. MOSA-Net uses cross-domain features (spectral and temporal features) and latent representations from a SSL model \cite{hubert} to predict objective quality and intelligibility scores simultaneously. As reported in \cite{mosa-net}, based on the cross-domain features and multi-task learning criterion, MOSA-Net can accurately predict objective quality (PESQ) \cite{rix2001perceptual} and intelligibility (STOI) \cite{ref_39} scores. In this paper, we propose an improved version of MOSA-Net, called the Multi-Target Speech Intelligibility Prediction Model (MTI-Net), for simultaneously predicting human and machine intelligibility scores. More specifically, machine intelligibility scores are word error rate (WER) scores from automatic speech recognition (ASR); and human intelligibility scores include: (1) subjective listening test results, and (2) objective evaluation scores (STOI in this study). MTI-Net is formed by a convolutional bidirectional long short-term memory (CNN-BLSTM) architecture with a multiplicative attention mechanism. To improve the prediction power, we further fine-tune the SSL model and use the fine-tuned embeddings (latent representations) as the input features for MTI-Net. Experimental results confirm that MTI-Net can predict human and machine speech intelligibility assessment scores well. With such multi-task learning criteria (simultaneous prediction of subjective listening test results, WER, and STOI), the prediction accuracy of individual outcomes can be improved. Furthermore, we observe that the fine-tuned SSL embeddings help MTI-Net achieve better results than SSL embeddings without fine-tuning.

The rest of this paper is organized as follows. Section II reviews related work. Section III presents the proposed MTI-Net. Section IV describes experimental setup and results. Finally, the conclusions and future work are presented in Section V. 

\section{Related Work}
\subsection{DL-based intelligibility prediction models}
Most DL-based speech assessment models aim to predict quality scores, and fewer work has focused on predicting intelligibility scores. In \cite{SIP_DL} and \cite{andersen2018nonintrusive}, intrusive and non-intrusive speech intelligibility prediction models are proposed; both employ a CNN model as the main model architecture and take compressed spectral features as input to predict measured intelligibility from multiple listening experiments. In \cite{ref_52} and \cite{mosa-net}, the CNN-BLSTM model architecture and attention mechanism are used to predict the objective intelligibility score (STOI). In \cite{dong2020towards}, the residuals of speech enhancement (SE) processed speech are incorporated to facilitate evaluating models to predict objective evaluation metrics (PESQ and ESTOI) without the need for a clean reference. Recently, a non-intrusive hearing aid speech assessment network (HASA-Net) \cite{chiang2021hasa} was proposed. Compared to other assessment models, HASA-Net takes the hearing loss pattern as an additional input to predict well-known hearing-aid evaluation metrics, namely the hearing aid speech quality index (HASQI) \cite{kates2014hearingb} and the hearing aid speech perception index (HASPI) \cite{kates2014hearinga}.

\subsection{Representations of SSL models}
Recently, the audio SSL model has attracted a lot of attention. Audio SSL models are trained from large-scale unlabeled data to learn representative embeddings \cite{liu2022audio}. It has been shown that SSL embeddings can be used as input features for various downstream tasks and yield promising performance \cite{nguyen2020zero, evain2021lebenchmark, yang2021superb, liu2022audio}. In \cite{huang2022investigating}, it was reported that fine-grained acoustic information may not be fully characterized by SSL embeddings. To further improve SSL embeddings to specific target tasks, some studies propose to take additional raw data or low-level features as input \cite{chen2021wavlm, mosa-net,nguyen2020investigating}.
 Meanwhile, fine-tuning SSL on the target downstream task also shows significant improvements. For examples, fine-tuning an SSL model improves three recognition tasks (speech emotion recognition, speaker verification, and spoken language understanding) \cite{wang2021fine}, end-to-end speech translation \cite{nguyen2020investigating}, and MOS prediction \cite{ssl-mos}. The above studies confirmed the effectiveness of combining cross-domain features with SSL embeddings and fine-tuning the SSL model to improve performance.

\section{Multi-Target Speech Intelligibility Prediction Model}
The overall architecture of MTI-Net is shown in Fig. 1. As shown in the figure, MTI-Net takes input from two branches. In the first branch, given a speech waveform \textbf{X}, the short-time Fourier transform (STFT) and learnable filter banks (LFB) are applied to generate two types of acoustic features, which are concatenated and sent to the convolutional layer. In the second branch, the speech waveform \textbf{X} is processed by an SSL model to obtain SSL embeddings. These two branches of features are concatenated and fed into a bidirectional layer and a fully connected layer. Finally, the output of the fully connected layer is mapped to three assessment metrics, namely the intelligibility, WER, and STOI scores. For each metric, an attention layer, a fully connected layer, and a global average pooling layer are applied to generate the final prediction output. To improve training stability, the objective function for training MTI-Net is a combined frame-level and utterance-level score, defined as follows:

\begin{equation}
\label{eq:loss}
   \small
    \begin{array}{c}
    O =  L_{I} + L_{W} + L_{S}
   \\ 
   L_{I}=\frac{1}{U}\sum\limits_{u=1}^U [(I_u-\hat{I}_u)^2+\frac{\alpha_I}{F_u}\sum\limits_{f=1}^{F_u}(I_u-\hat{i}_f)^2]
    \\
    L_{W}=\frac{1}{U}\sum\limits_{u=1}^U [(W_u-\hat{W}_u)^2+\frac{\alpha_W}{F_u}\sum\limits_{f=1}^{F_u}(W_u-\hat{w}_f)^2]
    \\
    L_{S}=\frac{1}{U}\sum\limits_{u=1}^U [(S_u-\hat{S}_u)^2+\frac{\alpha_S}{F_u}\sum\limits_{f=1}^{F_u}(S_u-\hat{s}_f)^2]   
    \end{array} 
\end{equation}
where $\{I_u,\hat{I}_u\}$, $\{W_u,\hat{W}_u\}$, and $\{S_u,\hat{S}_u\}$ are the true and predicted utterance-level scores of intelligibility, WER, and STOI, respectively; $U$ denotes the total number of training utterances; ${F}_u$ denotes the number of frames in the $u$-th training utterance; $\hat{i}_f$, $\hat{w}_f$, and $\hat{s}_f$ are the predicted frame-level scores of Intelligibility, WER, and STOI of the $f$-th frame, respectively; $\alpha_I$, $\alpha_W$, and $\alpha_S$ are the weights between utterance-level and frame-level losses.

\graphicspath{ {./images/} }
\begin{figure}[t]
\centering
\includegraphics[width=7cm]{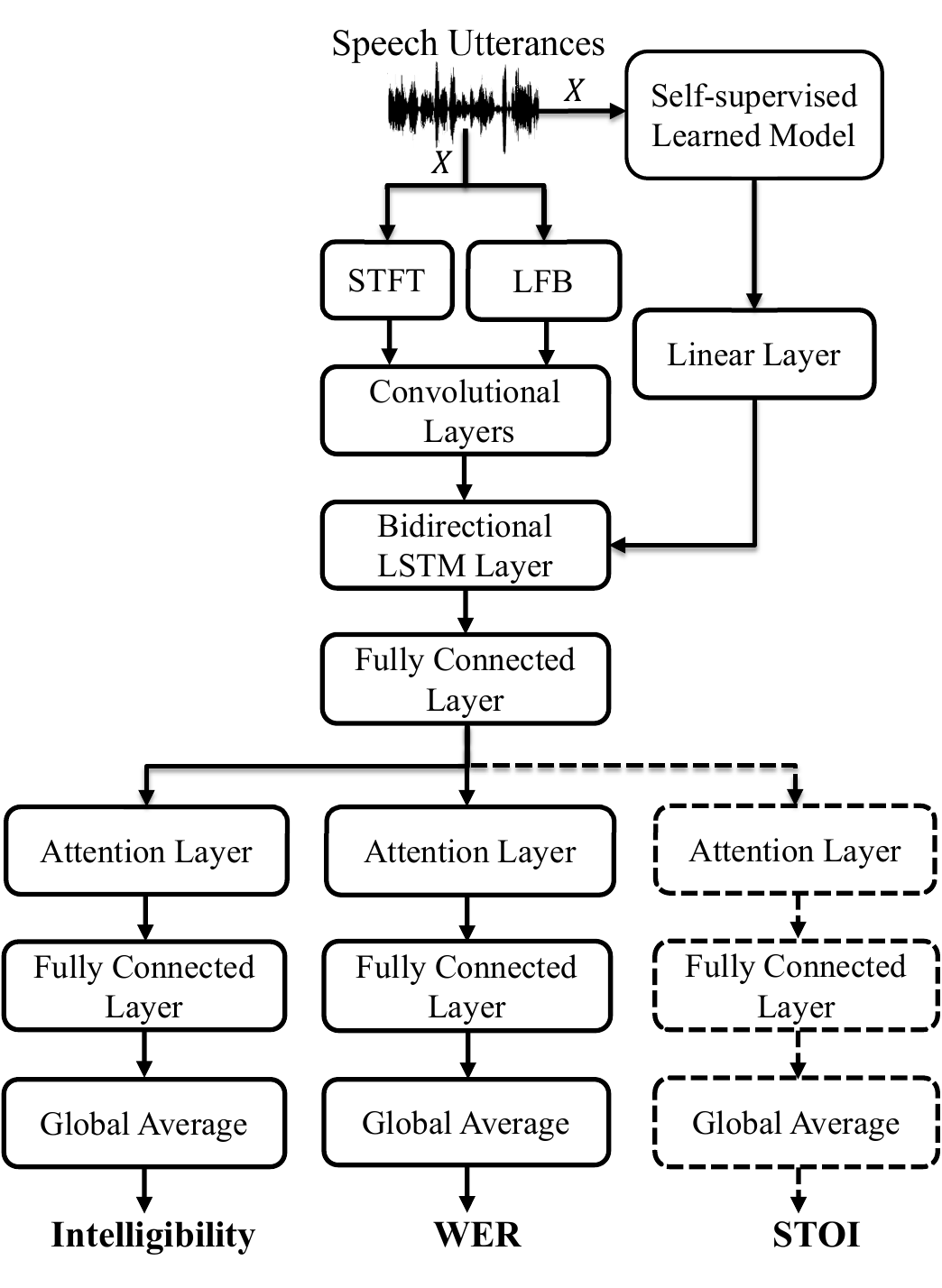} 
\caption{Architecture of the MTI-Net model.} 
\label{fig:MOSA-Net}
\end{figure}

To further improve the prediction accuracy of MTI-Net, we fine-tune the SSL model for the intelligibility prediction task. The process of fine-tuning the SSL model is defined as follows:
\begin{equation} \\
    \begin{array}{c}
    \textbf{Res} \ = \ SSL Model_{\theta} \ (\textbf{X}) 
    \\
    \textbf{Mean} \ = \ Mean Pooling\ (\textbf{Res})
    \\
    Estimated Score \ = \ Linear Layer_{\theta} \ (\textbf{Mean})
    \end{array} 
    \label{eq:QIA-SE}
\end{equation} 
where $\textbf{X}=[\textbf{x}_1,…,\textbf{x}_u…,\textbf{x}_U ]$ denotes the speech waveforms used for model training, \textbf{Res} denotes the embeddings from the SSL Model, \textbf{Mean} denotes the mean pooling of the embeddings. The parameter set $\theta$ of the SSL model is fine-tuned by minimizing the loss function based on the mean squared error (MSE), i.e., the first part of $L_{I}$ in Eq. (\ref{eq:loss}). Note that during the training of the MTI-Net with SSL features, the parameters of the pre-trained or fine-tuned SSL model are fixed. Therefore, the training objective function of the MTI-Net with fine-tuned SSL features is still Eq. (\ref{eq:loss}). 

%\begin{equation}
%\theta=\mathop{\arg\min}_{\theta}L(Label Score, Estimated Score),
%\end{equation}
%where \emph{Label Score} is the ground-truth scores of intelligibility, WER and STOI, and \emph{Estimated Score} denotes the predicted scores from MTI-Net.

In this study, we focus on the following two points: (1) As reported in \cite{ssl-mos}, fine-tuning the SSL model can already achieve satisfactory MOS prediction results. We intend to explore whether fine-tuning SSL embeddings in combination with low-level (but fine-grained) features can yield further improvements in predicting speech intelligibility. (2) We intend to investigate whether the multi-task learning criterion for predicting scorers of subjective listening tests, ASR results, and objective STOI scores can improve prediction performance.

\section{Experiments}
\subsection{Experimental setup}
We evaluated the proposed MTI-Net model on the Taiwan Mandarin Hearing In Noise test - Quality $\&$ Intelligibility (TMHINT-QI) \cite{TMINT-QI} dataset. The dataset includes clean, noisy, and enhanced utterances from five different SE systems, including Karhunen-Loeve transform (KLT) \cite{klt}, minimum-mean squared error (MMSE) \cite{mmse}, fully convolutional network (FCN) \cite{FCN}, deep denoising autoencoder (DDAE) \cite{DDAE}, and transformer-based SE \cite{Trans}). There are 226 subjects participated in the listening test \footnote{Written informed consent approved by the Academia Sinica Institutional Review Board for this study was obtained from each participant before conducting the experiment.}, and each listener was requested to rate the quality and answer what she/he heard for 108 utterances. For more details on the TMHINT-QI dataset, please see \cite{TMINT-QI}.
In this study, we selected the intelligibility score as one target. Another target, WER, was obtained by performing recognition on each utterance using the open-source Google ASR system \cite{ASR}. 
Each subjective intelligibility score ranges from zero to one and represents the percentage of correctly recognized characters. Similarly, each WER score ranges from zero to one, with lower scores indicating higher ASR accuracy. \par
To prepare the training set, we selected 15,000 utterances, each of which was evaluated by one listener. To prepare the test set, we selected 1,900 utterances, each of which was evaluated by 2 to 3 listeners, and the average score of these listeners was used as the ground-truth score for each utterance. Notably, the training and test utterances did not overlap. We used three evaluation metrics, namely MSE, linear correlation coefficient (LCC), and Spearman's rank correlation coefficient (SRCC) \cite{srcc} to evaluate the performance of MTI-Net. A lower MSE value indicates that the predicted scores are closer to the ground-truth scores (lower is better), while higher LCC and SRCC scores indicate a higher correlation between predicted scores and ground-truth scores (higher is better).

\subsection{MTI-Net with different features and targets}
In the first experiment, we aim to compare the performance of MTI-Net using different features and targets. We tested performance using power spectral features, features extracted by learnable filter banks, and the SSL embeddings, denoted as PS, LFB, and SSL, respectively. The HuBERT model \cite{hubert} was used to extract the SSL embeddings. The cross-domain features (hereafter referred to as CS) stand for the use of three features: PS, LFB, and SSL. For a fair comparison, the same model architecture (CNN-BLSTM with attention) \cite{mosa-net} was used to test the performance of different features. We further tested two prediction targets: one only predicting listening test results (denoted as I), and the other one predicting both listening test results and WER scores (denoted as I+W). The experimental results are summarized in Table 1. From the table, we first note that MTI-Net with cross-domain features can achieve the best performance for both intelligibility and WER score prediction, confirming the advantage of combining acoustic information from different features. Next, when comparing the results of single-target (I) and two-target (I+W), the (I+W) system always outperform its (I) counterpart. The results demonstrate that multi-task learning enables MTI-Net to achieve better speech intelligibility and WER predictions. In the following, we will adopt MTI-Net using cross-domain features and two-target (I+W) as the base system for further comparison. 

\begin{table}[t]
\caption{LCC, SRCC, and MSE results of MTI-Net using different features and targets. PS, LFB, and SSL, respectively, denote power spectral features, features extracted by learnable filter banks, and the SSL embeddings. CS denotes the cross-domain features. I and W denote the intelligibility and WER scores, respectively, which are the prediction targets of MTI-Net.}
\footnotesize
\begin{center}
 \begin{tabular}{c||c||c||c||c} 
 \hline
 \hline
 \textbf{Feature} &\textbf{Target} &\textbf{LCC} & \textbf{SRCC} & \textbf{MSE}  \\ [0.5ex] \cline{2-4}
 \hline\hline
  \multicolumn{5}{c} {Intelligibility Score Prediction} \\
 \hline
PS&I&0.543&0.483&0.034\\ \hline
PS&I+W&0.585&0.570&0.031\\ \hline
LFB&I&0.386&0.371&0.041\\ \hline
LFB&I+W&0.588&0.536&0.032\\ \hline
SSL&I&0.630&0.610&0.003\\ \hline
SSL&I+W&0.640&0.625&0.029\\ \hline
CS&I&0.731&0.685&0.022\\ \hline
CS&I+W&\textbf{0.761}&\textbf{0.708}&\textbf{0.020}\\ \hline
 \hline
  \multicolumn{5}{c} {WER Score Prediction} \\
 \hline
PS&W&0.661&0.644&0.057\\ \hline
PS&I+W&0.611&0.615&0.065\\ \hline
LFB&W&0.576&0.514&0.066 \\ \hline
LFB&I+W&0.583&0.542&0.066\\ \hline
SSL&W&0.734&0.725&0.047\\ \hline
SSL&I+W&0.690&0.687&0.052\\ \hline
CS&W&0.774&0.764&0.040\\ \hline
CS&I+W&\textbf{0.779}&\textbf{0.766}&\textbf{0.040}\\ \hline

 \hline

\end{tabular}
\end{center}
\end{table}

\subsection{MTI-Net with knowledge transfer (KT) and multi-task learning (MTL) methods}

%In this experiment, we aim to elaborate the generalization of MTI-Net from the perspective of performing model adaptation and expanding the ground-truth label. 
The main task of MTI-Net is to predict intelligibility and WER scores. We examine whether the STOI metric provides MTI-Net useful information for better prediction ability. Therefore, in this experiment, we investigate two methods for integrating STOI information into MTI-Net. The first method is based on knowledge transfer (KT), and the second method is based on multi-task learning (MTL). For the KT method, we adopted the pretrained network STOI-Net \cite{ref_52} as the seed model and fine-tuned the model to suit the prediction task of this study. For the MTL method, we simply used the STOI score as an additional target, as shown in Fig. 1. The best system in Table 1, i.e., CS(I+W), was used as the base model (denoted as Base in Table 2). The base models refined by the KT and MTL methods are denoted as KT and MTL, respectively, in Table 2. Table 2 lists the results of KT and MTL. The results of CS(I+W) in Table 1, are also listed and termed Base in Table 2 for comparison. As can been seen from Table 2, KT achieves better performance than Base in both intelligibility and WER prediction, confirming the advantages of the KT method. Furthermore, MTL achieves the best performance among the three methods, indicating that incorporating the STOI score at the target can effectively improve the prediction accuracy.  
%It indicates that by having better initialized parameter, it can improve the generalization of speech assessment model. 

\begin{table}[t]
\caption{LCC, SRCC, and MSE results of MTI-Net with knowledge transfer (KT) and multi-task learning (MTL).}
\footnotesize
\begin{center}
 \begin{tabular}{c||c||c||c||c} 
 \hline
 \hline
 \textbf{Methods} &\textbf{Target} &\textbf{LCC} & \textbf{SRCC} & \textbf{MSE}  \\ [0.5ex] \cline{2-4}
 \hline\hline
  \multicolumn{5}{c} {Intelligibility Score Prediction} \\
 \hline

Base&I+W&0.761&0.708&0.020\\ \hline
KT&I+W&0.781&0.713&0.019\\ \hline
MTL&I+W+S&\textbf{0.789}&\textbf{0.722}&\textbf{0.018}\\ \hline
 \hline
  \multicolumn{5}{c} {WER Score Prediction} \\
 \hline

Base&I+W&0.779&0.766&0.040\\ \hline
KT&I+W&0.798&0.780&0.036\\ \hline
MTL&I+W+S&\textbf{0.811}&\textbf{0.802}&\textbf{0.034}\\ \hline

 \hline

\end{tabular}
\end{center}
\end{table}

\subsection{MTI-Net with fine-tuning SSL embeddings}
Next, we examine the effectiveness of fine-tuning SSL embeddings in MTI-Net. Table 3 shows the prediction results of two MTI-Net systems: (1) MTI-Net: the best system from Table 2, which uses cross-domain features without fine-tuning SSL embeddings; and (2) MTI-Net(FT-SSL): MTI-Net using fine-tuned SSL embeddings in cross-domain features. The results of directly fine-tuning the SSL model to predict intelligibility, WER and STOI scored are also listed for comparison, denoted as FT-SSL in Table 3. We followed Eq. (2) and \cite{ssl-mos} and used the script\footnote{https://github.com/nii-yamagishilab/mos-finetune-ssl} to implement FT-SSL. Note that compared to the fine-tuning process of the SSL model described in Section III, a slight modification was applied to enable multi-task learning. Specifically, the mean pooling of the embeddings from the SSL model was fed to three linear layers corresponding to intelligibility, WER, and STOI, respectively, instead of one linear layer for predicting intelligibility. From Table 3, we can see that MTI-Net using pre-trained SSL embeddings (without fine-tuning the SSL model during MTI-Net training) can already achieve promising performance, comparable to FT-SSL in all evaluation metrics. However, MTI-Net(FT-SSL) yields the best performance, confirming the effectiveness of fine-tuning SSL embeddings during MTI-Net training.
\begin{table}[t]
\caption{LCC, SRCC, and MSE results of MTI-Net without and with fine-tuning SSL embeddings, termed MTI-Net and MTI-Net(FT-SSL), respectively.}
\footnotesize
\begin{center}
 \begin{tabular}{c||c||c||c||c} 
 \hline
 \hline
 \textbf{Systems} &\textbf{Target} &\textbf{LCC} & \textbf{SRCC} & \textbf{MSE}  \\ [0.5ex] \cline{2-4}
 \hline\hline
  \multicolumn{5}{c} {Intelligibility Score Prediction} \\
 \hline

FT-SSL &I+W+S&0.795&0.678&0.024\\ \hline
MTI-Net&I+W+S&0.789&0.722&0.018\\ \hline
MTI-Net(FT-SSL)&I+W+S&\textbf{0.823}&\textbf{0.735}&\textbf{0.017}\\ \hline
 \hline
  \multicolumn{5}{c} {WER Score Prediction} \\
 \hline

FT-SSL&I+W+S&0.824&0.815&0.036\\ \hline
MTI-Net&I+W+S&0.811&0.802&0.034\\ \hline
MTI-Net(FT-SSL)&I+W+S&\textbf{0.834}&\textbf{0.822}&\textbf{0.031}\\ \hline

 \hline

\end{tabular}
\end{center}
\end{table}

%\subsection{Comparison of MTI-Net with related approaches}
%Finally, we compare the performance of MTI-Net with another multi-task learning based speech intelligibility's prediction model namely, inQSS \cite{TMINT-QI}, and MOSA-Net \cite{mosa-net}\footnote{https://github.com/dhimasryan/MOSA-Net-Cross-Domain}. Both of inQSS and MOSA-Net were originally trained on TMHINT-QI dataset, where the subjective quality and intelligibility were selected as the ground truth score to train the model. In detail, inQSS used spectral features and scattering coefficient as the input features to train CNN-BLSTM model. For MOSA-Net, the cross-domain features were used as the input features to train CNN-BLSTM with attention model. Experimental result indicates that MTI-Net can achieve better performance than inQSS and MOSA-Net in all evaluation metrics. Interestingly, by selecting more correlated additional ground truth labels while performing multi-task learning, it can significantly boost the prediction performance.
We also present the scatter plots of the predicted targets of MTI-Net(FT-SSL), MTI-Net, and FT-SSL in Fig. 2. The figure again confirms that MTI-Net(FT-SSL) can achieve more accurate intelligibility and WER predictions than MTI-Net and FT-SSL. More specifically, the points of MTI-Net(FT-SSL) are more densely distributed on the diagonal than those of MTI-Net and FT-SSL. The results in Tables 1-3 and Fig. 2 confirm the benefits of fine-tuning SSL embeddings, using cross-domain features, and the multi-task learning criterion of MTI-Net.

\graphicspath{ {./images/} }
\begin{figure}[t]
\centering
\includegraphics[width=8cm]{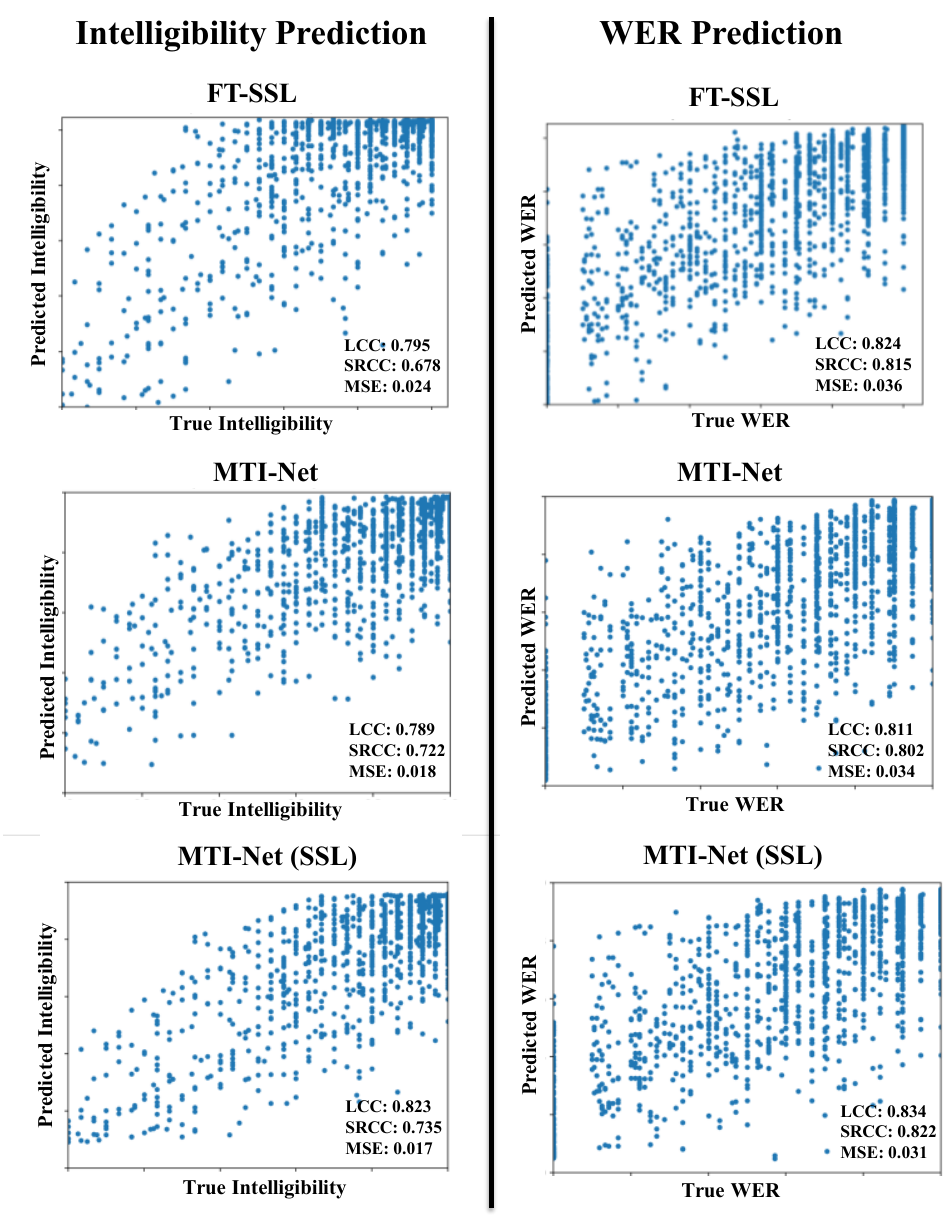} 
\caption{Scatter plots of three speech intelligibility prediction models, including FT-SSL \cite{ssl-mos}, MTI-Net, and MTI-Net(FT-SSL).} 
\label{fig:Scatter}
\end{figure}

\section{Conclusions}
In this paper, we have proposed MTI-Net that aims to predict human speech intelligibility and machine WER. MTI-Net adopts cross-domain features and a CNN-BLSTM model architecture with attention mechanism, and is trained by a multi-task learning criterion to predict both subjective listening test and WER scores simultaneously. In experiments, we first demonstrated the effectiveness of combining low-level (and fine-grained) features with SSL embeddings. Next, we confirmed the advantages of multi-task learning. Finally, we verified the positive results of fine-tuning SSL embeddings. To our knowledge, this is the first work that aimed to directly predict WER with high correlation scores. Our experimental results also confirmed that by combining information from subjective listening test and WER scores, along with objective STOI metrics, MTI-Net can more accurately predict intelligibility and WER scores. In the future, we will investigate using MTI-Net as a learnable objective function to guide SE model training.       

\bibliographystyle{IEEEtran}

\bibliography{mybib}

% \begin{thebibliography}{9}
% \bibitem[1]{Davis80-COP}
%   S.\ B.\ Davis and P.\ Mermelstein,
%   ``Comparison of parametric representation for monosyllabic word recognition in continuously spoken sentences,''
%   \textit{IEEE Transactions on Acoustics, Speech and Signal Processing}, vol.~28, no.~4, pp.~357--366, 1980.
% \bibitem[2]{Rabiner89-ATO}
%   L.\ R.\ Rabiner,
%   ``A tutorial on hidden Markov models and selected applications in speech recognition,''
%   \textit{Proceedings of the IEEE}, vol.~77, no.~2, pp.~257-286, 1989.
% \bibitem[3]{Hastie09-TEO}
%   T.\ Hastie, R.\ Tibshirani, and J.\ Friedman,
%   \textit{The Elements of Statistical Learning -- Data Mining, Inference, and Prediction}.
%   New York: Springer, 2009.
% \bibitem[4]{YourName17-XXX}
%   F.\ Lastname1, F.\ Lastname2, and F.\ Lastname3,
%   ``Title of your INTERSPEECH 2021 publication,''
%   in \textit{Interspeech 2021 -- 20\textsuperscript{th} Annual Conference of the International Speech Communication Association, September 15-19, Graz, Austria, Proceedings, Proceedings}, 2020, pp.~100--104.
% \end{thebibliography}

\end{document}